\documentclass{article}
\usepackage{graphicx}    
\usepackage{booktabs}
\usepackage{amsmath}
\usepackage{amssymb}     
\usepackage[margin=1in]{geometry}
\usepackage{natbib}
\usepackage{hyperref}
\usepackage{doi}
\usepackage[mathlines]{lineno}

\title{Heavy-Tailed Dispersal Kernels from Stopped Subdiffusive Fractional Brownian Motion}
\date{}

\author{Luis F. Gordillo\\ 
{\small Department of Mathematics and Statistics, Utah State University, Logan, UT}\\
Priscilla E. Greenwood\\ 
{\small Department of Mathematics, University of British Columbia, Vancouver, BC}}

\begin{document}

\maketitle
\begin{abstract}
    Subdiffusive fractional Brownian motions produce localized aggregation when  particles are stopped at exponentially distributed times. In  applications where clumping and long-distance dispersal events are observed simultaneously, such as in some instances of seed dispersal, this model fails to describe the tails of the  data.  The resulting redistribution kernel has only an exponentially decaying tail, whereas a heavier tail is needed for modeling the long-distance dispersal observed.
    Here we propose a  model in which subdiffusive particles stop at exponentially distributed times, but with a rate parameter that is Gamma distributed. This heterogeneity in stopping rates causes the density of final radial positions to have a heavy-tailed distribution. 
    Our model retains the strong localized clumping characteristic of subdiffusive fractional Brownian motion while simultaneously generating the heavy tails required for realistic long-distance dispersal.
\end{abstract}

\section{Introduction}
The paper (\cite{BK53}) is one of the first applied contributions to the study of  stopped Brownian motions. There, the spatial dispersal of \textit{Trichostrongylus retortaeformis} larvae (a nematode parasite) is modeled as an isotropic Brownian motion in the plane, starting at the origin. In the model, the larvae move until they settle on a grass blade, and then they are presumed to be ingested by a grazing host in the next step of their life cycle. The settling process is modeled as occurring at a constant rate, independent of position and time, which is equivalent to stopping the Brownian motion at an independent exponentially distributed random time $\tau$ with a parameter $\lambda > 0$. 
Broadbent and Kendall derived the radial probability density function of the larvae’s final positions. This density depends on the diffusion coefficient of the motion and on the stopping-rate parameter, and also involves a modified Bessel function of the second kind. The resulting redistribution kernel has a tail that decays exponentially, implying that the positions of settlement of individuals are localized around the initial point and long-distance dispersal of individuals is limited. 

However, a substantial body of evidence across many species demonstrates that long-distance dispersal events do occur and play a crucial role in the colonization of remote habitats, the spread of invasive species, and species coexistence (\cite{johnson2026modeling,Nathan2006LongDistance, Trakhtenbrot2005LDD}). 
A key question is how to model the occurrence of long-distance dispersal in a realistic manner. This problem has been studied extensively over the past decades (\cite{lewis2016mathematics, Nathan2008MovementEcology}), highlighting the challenges posed by individual movement complexity and natural unpredictability. 
Despite this progress, the literature remains relatively limited in terms of mechanistic models that explain the emergence of heavy-tailed dispersal kernels in long-distance dispersal, rather than simply fitting fat-tailed distributions phenomenologically. 
Here, we propose a mechanistic framework that naturally accounts for the possible emergence of heavy-tailed dispersal kernels in real scenarios.
    
Suppose that due to variability, either biological or in transport, the individuals in the population have exponential stopping times but with a different parameter for each individual. Then a distribution of the stopping positions will be significantly broader. 
Specifically, (\cite{gordillo2025parameter}) shows that the distribution of the final positions of stopped Brownian motions is heavy-tailed when the parameter $\lambda$ follows a Gamma distribution. Thus, introducing randomness in the stopping rate, i.e. heterogeneity in stopping rates across individuals, allows for long-distance dispersal in the model.
In this paper we prove that this phenomenon extends even to the case when each individual's movement is modeled via subdiffusive fractional Brownian motion (fBm), characterized by having a Hurst parameter $H < 1/2$. The mean square displacement of individuals grows as $t^{2H}$ as time $t$ increases. Subdiffusive fBm spreads more slowly and with more local clumping than standard Brownian motion ($H = 1/2$). We give an analytic approximation argument and verify by simulation that the radial density $g(r)$ of stopped positions satisfies $g(r) \sim C r^{-1 - a/H} $ as $ r \to \infty $, $C$ and $a>0$, producing a power-law tail with index $a/H$. Here $a$ is the scale parameter in $\mathrm{Gamma}(a,b)$, the distribution of the parameter $\lambda$ in the exponentially distributed stopping times.

\section{Subdiffusive fBm}
In this Section we provide the necessary background from the theory of fBm. More complete treatments of fBm can be found in (\cite{biagini2008}) Chapter 1, or (\cite{SamorodnitskyTaqqu1994}), Chapter 7.
Let us denote by $B_H(t)$ a fBm with Hurst exponent $H \in (0,1)$. It is a self-similar, centered Gaussian process with stationary increments that starts at $B_H(0) = 0$ and has covariance function
\begin{equation}\label{eq: covariance}
\mathbb{E}[B_H(t) B_H(s)] = \frac{\sigma^2}{2} \left( t^{2H} + s^{2H} - |t - s|^{2H} \right),
\end{equation}
for $t,s\geq 0$, where the constant $\sigma^2$ depends on the physical applications. 
Long-range temporal correlations, i.e. long memory or long-range dependence, depend on whether $H > 1/2$, $H = 1/2$, or $H < 1/2$.
Expression \eqref{eq: covariance} recovers the covariance of the standard Brownian motion when $H = 1/2$.

\subsection{Behavior of correlations between increments}
Consider the increment process of $B_H$, known as fractional Gaussian noise, evaluated at discrete integer times for notational simplicity
\[X_k = B_H(k+1) - B_H(k), \quad k = 0, 1, 2, \dots\]
The continuous-time analogue follows identically. 
The covariance between increments disjoint increments with lag $n \geq 1$ is obtained by using \eqref{eq: covariance},
\begin{align*}
    \mathbb{E}[X_k X_{k+n}] &= \mathbb{E}\bigl[ \bigl(B_H(k+1)-B_H(k)\bigr) \bigl(B_H(k+n+1)-B_H(k+n)\bigr) \bigr]\\
    &= \frac{\sigma^2}{2}\big[(n+1)^{2H}+(n-1)^{2H}-n^{2H}\big]
\end{align*}
As the size of the lag increases, $n \to \infty$, this covariance admits the asymptotic expansion
$$\mathbb{E}[X_k X_{k+n}] \sim \sigma^2 H(2H-1) \, n^{2H-2}.$$
For the subdiffusive case, $0 < H < 1/2$, the factor $H(2H-1) < 0$,  since $H<1/2$, so the covariance is negative for large lags. This reflects antipersistence behavior: distant increments tend to have opposite signs, which prevents fast escape from the origin and produces the subdiffusive spreading, even though the single-time marginal distribution remains Gaussian.

The ``long-range” designation comes from the dependence between the increments $X_k$ and $X_{k+n}$ remaining ``geometric" for arbitrarily large $n$. The process ``remembers” its past at a geometrically decreasing rate over all time horizons.

\subsection{The pdf of subdiffusive fBm}
The radial probability density function of the distance $r = \sqrt{x^2 + y^2}$ from the origin in a 2-dimensional subdiffusive fBm (with Hurst exponent $H < 1/2$) follows directly from the isotropic 2-dimensional pdf,
\[
P(x, y, t) =  \frac{1}{2\pi \sigma^2 t^{2H}} \exp\left( -\frac{x^2+y^2}{2 \sigma^2 t^{2H}} \right),
\]
where \( \sigma^2 > 0 \) is the variance factor.
The corresponding radial pdf \( \phi(r, t) \) 
is
\begin{equation}\label{eq: radial pdf}
\phi(r, t) =\frac{r}{\sigma^2 t^{2H}} \exp\left( -\frac{r^2}{2 \sigma^2 t^{2H}} \right),
\end{equation}
which is a Rayleigh distribution with time-dependent scale parameter $s(t) = \sigma t^H$.
The marginal distribution at any fixed time is Gaussian, even though the underlying process has long-range negative temporal correlations (encoded in the covariance, not in the single-time marginal).

\section{Stopping of subdiffusive fBm}
In contrast to (\cite{BK53}), who take their diffusion to be Brownian motion, we assume that each individual moves in the plane with subdiffusive fBm and stops at a time $\tau$. If the stopping time is exponentially distributed with parameter $\lambda>0$, constant, the radial distribution for the stopping locations, $g(r)$, is given by
\begin{equation}
    g(r)=\frac{\lambda}{2\pi\sigma^2} \int_0^\infty \frac{1}{t^{2H}} \exp\left( -\frac{r^2}{2\sigma^2 t^{2H}} - \lambda t \right) \, dt.
\end{equation}
Suppose now that $\lambda$ is not the same for all individuals but has a probability distribution  over $(0,\infty)$ for each individual, i.e. $\lambda$ is a random variable with probability density $f_\lambda (s)$ having support $(0,\infty)$. The distribution of the stopping time is
\begin{equation}\label{eq: stopping time}
\Theta (t)=P(\tau \leq t)=\int_0^\infty P(\tau \leq t|\lambda=s)f_\lambda(s)ds 
= \int_0^\infty (1-e^{-st})f_\lambda(s)ds .
\end{equation}
By combining \eqref{eq: radial pdf} and \eqref{eq: stopping time} we write the radial probability density, $g(r)$, of the final location of a particle as
\begin{equation}\label{eq: general g(r)}
g(r) = \int_0^\infty \phi(r,t)\cdot\frac{d\Theta}{dt}\,dt.
\end{equation}
The primary aim of this Section is to derive the asymptotic rate of the tail of $g(r)$ as $r\rightarrow\infty$, where $\phi(r,t)$ is given by \eqref{eq: radial pdf}. 

Equation \eqref{eq: general g(r)}, a ``mixture" model, represents the radial distribution of particles where the transport of each individual particle may be subject to different modes, such as wind,
attachment to animal fur or ingestion by animals, that produce different particular stopping time distributions. 
In our development \eqref{eq: stopping time} and \eqref{eq: general g(r)}, each particle has a stopping time that is exponentially distributed
with parameter $\lambda$ chosen from a distribution. The distribution we choose here will be a
Gamma distribution, i.e. $\lambda \sim \mathrm{Gamma}(a,b)$. The parameters $a$ and $b$ control the shape and the scale of the distribution, and can be chosen to model a very diverse set of transport characteristics.
Equation \eqref{eq: stopping time} yields
\[
    \Theta(t)=\int_0^\infty \left( 1-e^{-st}\right)\cdot\frac{b^a}{\Gamma(a)}s^{a-1}e^{-bs} ds
    = 1-\frac{b^a}{\Gamma(a)}\int_0^\infty s^{a-1} e^{-(t+b)s}ds
    = 1- \frac{b^a}{(t+b)^a}.
\]
and
\begin{equation}\label{eq: Pareto}
    \frac{d\Theta}{dt} = \frac{a b^a}{(t + b)^{(a+1)}}, \quad t > 0,
\end{equation}
which we recognize as a shifted Pareto, or Lomax, distribution (\cite{johnson1994vol1}).
It follows from \eqref{eq: general g(r)} that the radial probability density $g(r)$ is
\begin{equation}\label{eq: g(r)}
    g(r) = \frac{r a b^a}{\sigma^2} \int_0^\infty  \frac{1}{t^{2H}}\exp\left( -\frac{r^2}{2 \sigma^2 t^{2H}} \right)\,\frac{1}{(t + b)^{a+1}} \, dt. 
\end{equation}
For the case \(H = 1/2\) and with the change of variables $x=1/t$ the latter can be expressed in closed form by using a confluent hypergeometric function  (\cite{gordillo2025parameter}), while for \(H \neq 1/2\) it does not admit a simple analytical form. In this paper we are interested in the tail behavior of \(g(r)\), which can be analyzed for the subdiffusive case, $H<1/2$, by using the following arguments.

\subsection{Tail analysis}
Let us analyze the tail behavior of the density \eqref{eq: g(r)}. The exponential term in the integrand of \eqref{eq: g(r)}, $\exp(-c / t^{2H})$, $c=r^2/2 \sigma^2$, is close to zero when $t$ is small and it approaches 1 when $t$ is sufficiently large. This behavior indicates that for small to moderate $t$ the contribution of the exponential term to the integral is very small, i.e. when $c/t^{2H} \gg 1$, or $t \ll c^{1/2H}$, but increases significantly once $t$ is large enough, i.e. when $c/t^{2H} \ll 1$, or $t \gg c^{1/2H}$.
Hence, 
$$t\sim c^{1/(2H)} \propto r^{2/(2H)} = r^{1/H}$$
defines the characteristic scale, i.e. the time threshold where the integrand contribution starts to be noticeable.
Since for subdiffusion $1/H>2$ holds, $r^{1/H}$ grows fast when $r\rightarrow\infty$. 
This means that the main contribution to the value of the integral comes from relatively large values of $t$ in the large-$r$ limit. In addition, $(t + b)^{-(a+1)} \approx t^{-(a+1)}$ becomes a good approximation when $t \gg b$.
Therefore, in order to evaluate the tail behavior of $g(r)$ in \eqref{eq: general g(r)} we consider the approximation to \eqref{eq: g(r)}
\begin{equation}\label{eq: g(r) approximation} 
    g(r)\sim\frac{rab^a}{\sigma^2}\int_0^\infty \frac{1}{t^{2H+a+1}} \exp\left( -\frac{r^2}{2 \sigma^2 t^{2H}} \right) \, dt, 
\end{equation}
when $r\rightarrow\infty$.  We can evaluate the integral  \eqref{eq: g(r) approximation} by using the change of variable $v=c/t^{2H}$,
\begin{align*}
\int_0^\infty \frac{1}{t^{2H+a+1}} \exp\left( -\frac{c}{ t^{2H}} \right) \, dt
&=\int_0^\infty \left( c^{1/2H} v^{-1/2H} \right)^{-(2H + a + 1)} e^{-v} \cdot \frac{1}{2H} c^{1/2H} v^{-(1 + 2H)/2H} \, dv\\[2ex]
&= \frac{1}{2H} c^{-1 - a/2H} \int_0^\infty v^{a/2H} e^{-v} \, dv \\
&= \frac{1}{2H} c^{-1 - a/2H} \,\Gamma\left( \frac{a}{2H} + 1 \right).
\end{align*}
Using this in the asymptotic expression for $g(r)$ gives
\begin{equation}\label{eq: final radial density}
    g(r) \sim \frac{r a b^a}{\sigma^2} \cdot \frac{1}{2H} \cdot r^{-2 - a/H} (2 \sigma^2)^{1 + a/(2H)} \Gamma\left(1 + \frac{a}{2H}\right) 
    = \frac{a b^a 2^{a/2H}}{H} \,\Gamma\left(1 + \frac{a}{2H}\right) \sigma^{a/H} \, r^{-1 - a/H},
\end{equation}
for $r\to\infty$. 
Let $\theta$ be an arbitrary but fixed positive value. If one compares the approximation of $g(r)$ obtained in \eqref{eq: final radial density} with the exponential decay $e^{-\theta r}$ as $r\to\infty$, one has
\[\frac{g(r)}{e^{-\theta r}} \sim C \frac{e^{\theta r}}{r^{1+a/H}}\to\infty, \qquad C=\frac{a b^a 2^{a/2H}}{H} \,\Gamma\left(1 + \frac{a}{2H}\right) \sigma^{a/H}.\]
This shows that the probability in the tail of $g(r)$ decreases at a rate slower than that of any exponential and therefore the distribution of radial distances is heavy-tailed. Alternatively, let $R$ be the radial distance of a stopped point. Integrating the tail approximation \eqref{eq: final radial density} gives
\begin{equation}\label{eq: regular variation}
P(R>r)\sim\int_r^\infty \frac{C}{s^{1+a/H}}\,ds =\frac{HC}{ar^{a/H}},
\end{equation}
which also decreases slower than $e^{-\theta r}$ for arbitrary $\theta>0$. Hence, the distribution of radial distances is heavy-tailed.

The heavy-tailed shifted Pareto density of stopping times \eqref{eq: Pareto} produces a slow decay via the time-weighting factor $(t+b)^{-(a+1)}$. Looking at \eqref{eq: g(r)} we can interpret the interaction between time and space probability densities as follows. Some of the moving particles have unusually long stopping times, i.e. large $t$, from the shifted Pareto distribution.  These particles can reach large distances because the Gaussian-like subdiffusive spreading is broad enough ($\sim t^H$) to cover large $r$. The probability of this combination of events is expressed by the product of the densities in the integrand in \eqref{eq: g(r)}. Hence, for large $r$, the tail of $g(r)$ is controlled by the power-law tail of the stopping-time density.

\subsection{Illustration by simulations}
\begin{figure}
    \centering
    \includegraphics[scale = 0.43]{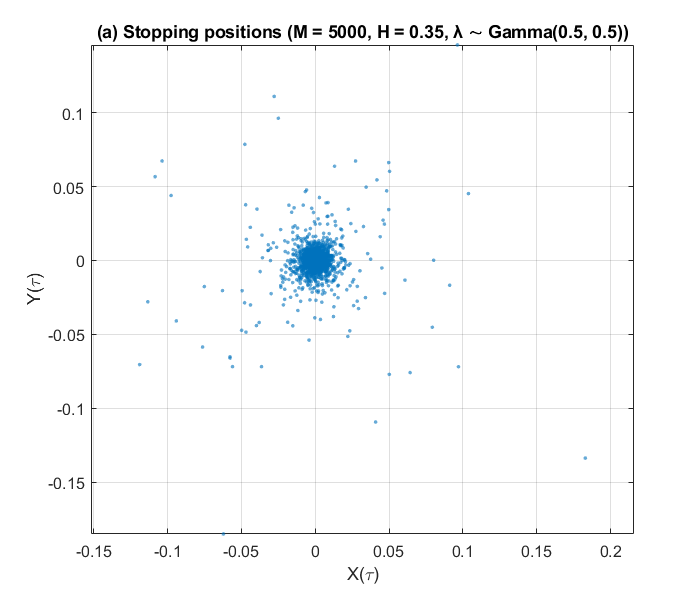}
    \includegraphics[scale = 0.43]{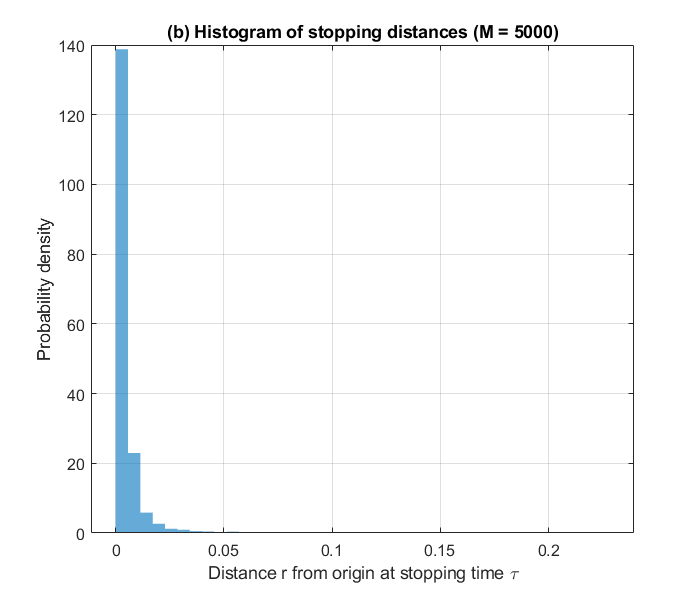}
    \includegraphics[scale = 0.43]{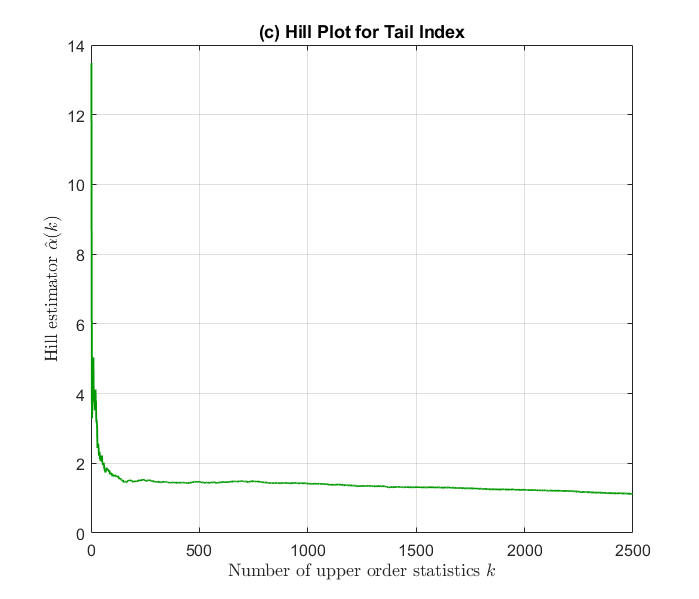}
    \includegraphics[scale = 0.43]{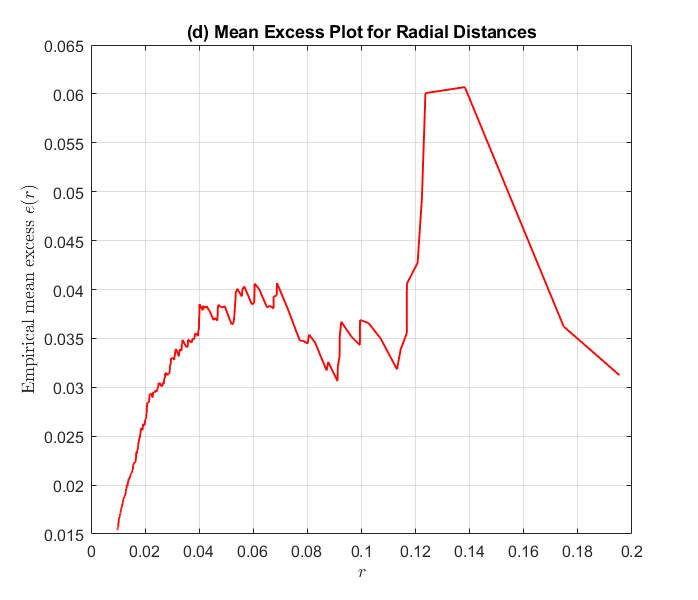}
    \includegraphics[scale = 0.43]{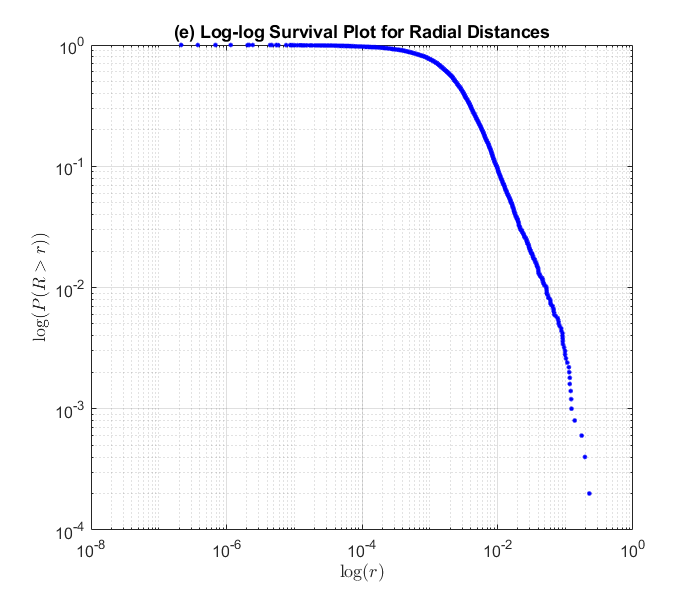}
    \caption{Radial distribution of stopped paths of a subdiffusive fBm with  randomized stopping rate $\lambda\sim \text{Gamma}(a,b)$.  (a) Stopped positions, (b) Histogram of stopping distances. The diagnostic plots:  Hill (c), Mean-excess (d) and Log-log survival (e), consistently support the presence of a heavy tail in the distribution of radial distances, see text.}
    \label{fig: Figure 1}
\end{figure}

Here we complement the tail analysis by simulating 5000 paths of a fBm with $H=0.35$, $\sigma=1$, and exponentially distributed stopping times, where the parameter $\lambda$ is randomized according to $\lambda\sim\mathrm{Gamma}(0.5,0.5)$.
The scatter plot of final positions, Figure \eqref{fig: Figure 1}a shows a dense central cloud and outliers. The histogram of stopping distances, Figure \eqref{fig: Figure 1}b, is right-skewed, with the bulk of probability mass near the origin and with a tail that, according to the scatter plot, may extend beyond \(r \approx 0.2\). 
The statistical diagnostic plots for heavy tails presented in Figures \eqref{fig: Figure 1}c, \eqref{fig: Figure 1}d and \eqref{fig: Figure 1}e support the presence of a heavy tail in the radial distribution of the stopping positions (\cite{nair2022fundamentals}):

\begin{enumerate}
    \item[i.] \textbf{Figure \eqref{fig: Figure 1}c: Hill plot} (tail-index estimator $\hat{\alpha}(k)$).  The Hill estimator is a nonparametric estimator used to estimate the tail index of a heavy-tailed distribution, based on a descendingly ordered sample $R_{(1)},...,R_{(n)}$, where regular variation of the tail is assumed, i.e. $P(R>r)\sim 1/r^\alpha$, $\alpha>0$. We argued in \eqref{eq: regular variation} that regular variation holds under our assumptions, where $\alpha=a/H$. The Hill estimator is defined by the upper order statistics,
    \[\hat{\alpha}(k)=\frac{1}{\frac{1}{k}\sum_{i=1}^k\log\left( \frac{R_{(i)}}{R_{(k)}}\right)},\quad k=1,...,n.\]
    
    In our example, the Hill plot starts at high values (around 14) and decreases rapidly, stabilizing to around 1.5 as the number of upper-order statistics $k$ increases to 1000, consistent with the theoretical value $a/H=0.5/0.35\approx1.43$.  Approximation of the Hill plot values near a finite positive constant for relatively large $k$ is a standard indicator of regularly varying tails with finite tail index $\alpha > 0$. Light-tailed distributions, e.g., Gaussian or exponential, would drive the estimator toward zero. This stabilization provides direct graphical evidence of a heavy tail.

    \item[ii.] \textbf{Figure \eqref{fig: Figure 1}d: Mean-excess plot}.  
  The empirical mean-excess function $e(r)$, defined as
  \[ e(r)=\mathbb{E}(R-r|R>r),\]
  is expected to grow unboundedly for heavy-tailed distributions. In contrast, for light-tailed distributions $e(r)$ is expected to decrease. In practice, the empirical mean-excess, defined by
  \[e_n(r)=\frac{\sum_{i=1}^n(R_i-r)I(R_i>r)}{\sum_{i=1}^nI(R_i>r)},\] 
  where $I$ is the indicator function, is used as a proxi for $e(r)$. If we look at values of $r$ below 0.04 then the empirical mean-excess plot shows a regular increasing trend. After that point, irregularities attributable to finite-sample effects in the tail make the test unusable. 
  The observed upward trend for $r$ below 0.04 constitutes strong supporting evidence of a heavy tail.

    \item[iii.] \textbf{Figure \eqref{fig: Figure 1}e: Log-log survival plot}.  
  The log of the empirical survival function remains near one for small $\log r$, then exhibits a region of relatively slow, approximately linear decay over several orders of magnitude, the hallmark of power-law tails, before the expected finite-sample cutoff.
\end{enumerate}

\begin{figure}
    \centering
    \includegraphics[scale = 0.43]{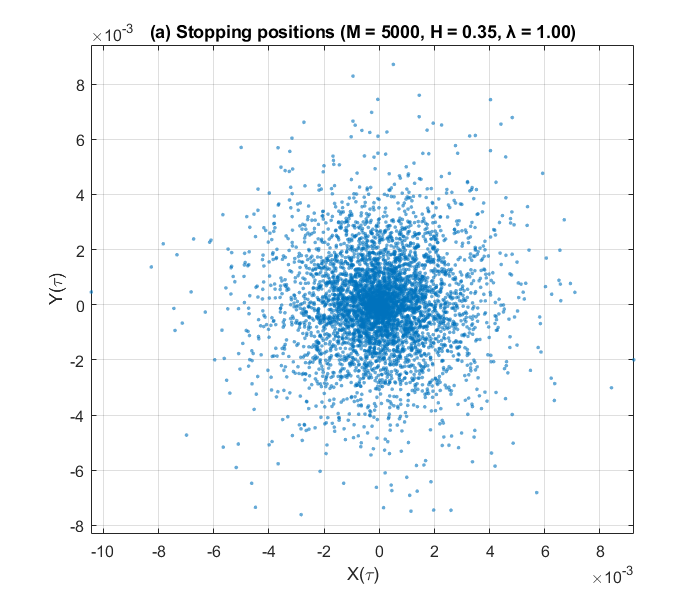}
    \includegraphics[scale = 0.43]{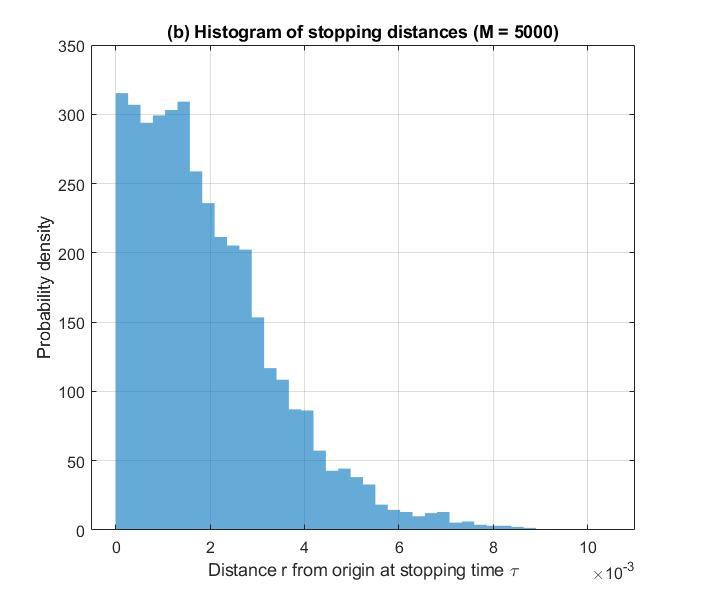}
    \includegraphics[scale = 0.43]{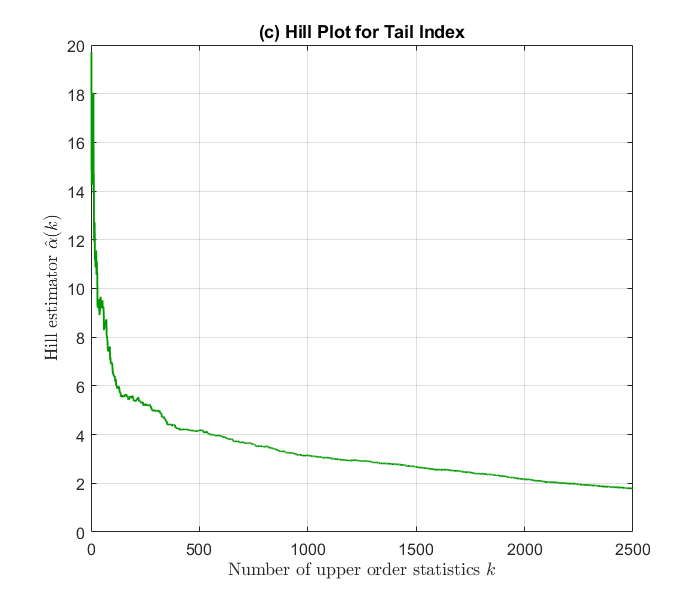}
    \includegraphics[scale = 0.43]{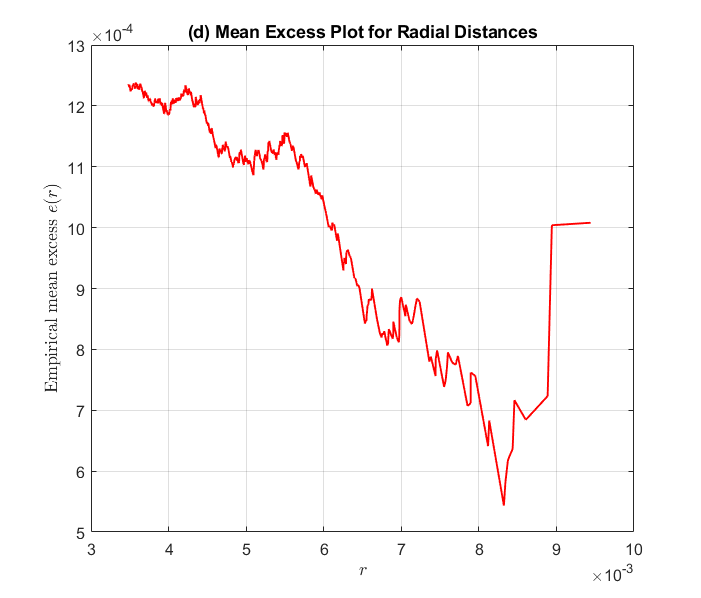}
    \includegraphics[scale = 0.43]{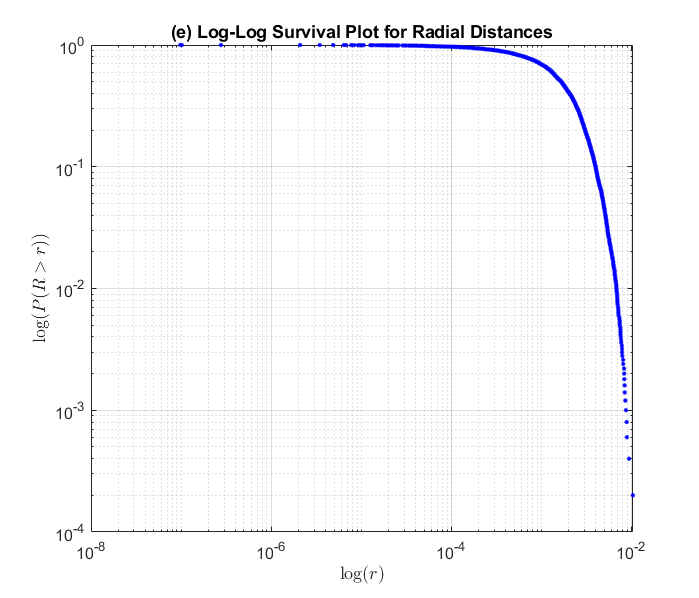}
    \caption{Same plot as Figure \eqref{fig: Figure 1} but with constant stopping rate $\lambda=1$.   Notice the difference in scale from Figure \eqref{fig: Figure 1}, see also Figure \eqref{fig: Figure 3}. The diagnostic plots  (c), (d) and (e) support the absence of a heavy tail.}
    \label{fig: Figure 2}
\end{figure}

\begin{figure}
    \centering
    \includegraphics[width=0.5\linewidth]{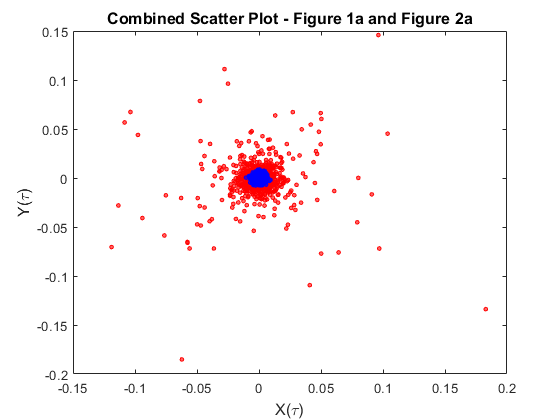}
    \caption{Scatter plots in Figure \eqref{fig: Figure 1}a (red) and Figure \eqref{fig: Figure 2}a (blue) superposed for comparison.}
    \label{fig: Figure 3}
\end{figure}
In Figure \eqref{fig: Figure 2} we repeat the simulations in Figure \eqref{fig: Figure 1} but with constant $\lambda=1$. The diagnostic plots confirm that the tail of the distribution of $r$ is not heavy.

\section{Parameter randomization models heterogeneity}
There is a significant value in recognizing how a Gamma distribution randomization of the exponential rate parameter yields a shifted Pareto distribution for the stopping times. It is a mathematically smooth way to model heterogeneity of the duration of motion of particles. This type of formulation may also represent heterogeneity in space, leading to additional realistic models involving subdiffusion.  

When, as here, the rate \(\lambda\) for the exponential stopping time $\tau\sim \mathcal{E}(\lambda)$ is itself drawn from a Gamma distribution, $\lambda \sim \mathrm{Gamma}(a, b)$, the marginal distribution of $\tau$ is a shifted-Pareto distribution with density $a b^a/(t + b)^{a+1}$ for $t > 0$. This is a classic compound distribution result, useful because it transforms a light-tailed (exponential) process into a heavy-tailed one without the need to postulate heavy tails directly. The shifted Pareto adds a scale parameter $b$ that removes the singularity at $t=0$, making the model physically plausible in systems where very short stopping times are unlikely.

In our simulations above, each individual particle has a distinct exponential distribution for the stopping time, possibly due to physical or phenotypical characteristics, such as size or mass. Assigning to each particle a rate parameter $\lambda$ independently drawn from, for example, a Gamma distribution is an approach for modeling particle heterogeneity.

\subsection{Example: Stratified dispersal}
In invasion biology, the term \textit{stratified dispersal} refers to the spread of species through two distinct spatial scales: localized, short-distance dispersal around established foci and infrequent long-distance dispersal events that generate new satellite populations (\cite{lewis2016mathematics, Shigesada1995Stratified}). Cheatgrass (\textit{Bromus tectorum}), a highly successful invasive annual grass in the western United States, exemplifies this pattern. Most cheatgrass seeds disperse over short distances via gravity, wind gusts, or attachment to small animals, resulting in dense local clumping near introduction points or disturbed sites. Concurrently, rare long-distance dispersal events, often human- or large animal-mediated, establish isolated founder populations that serve as nuclei for further expansion (\cite{Molvar2024Cheatgrass,lewis2016mathematics,Zouhar2003BromusTectorum}).
These characteritics align with those produced by our model, i.e. localized clumping and long range spread.

\begin{figure}
    \centering
    \includegraphics[scale=0.5]{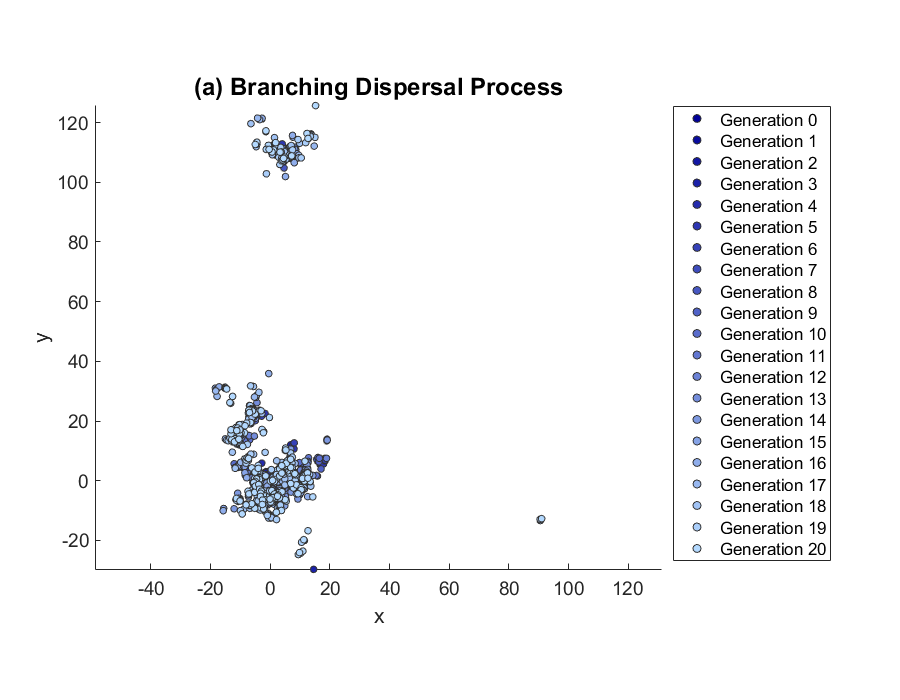}
    \includegraphics[scale=0.5]{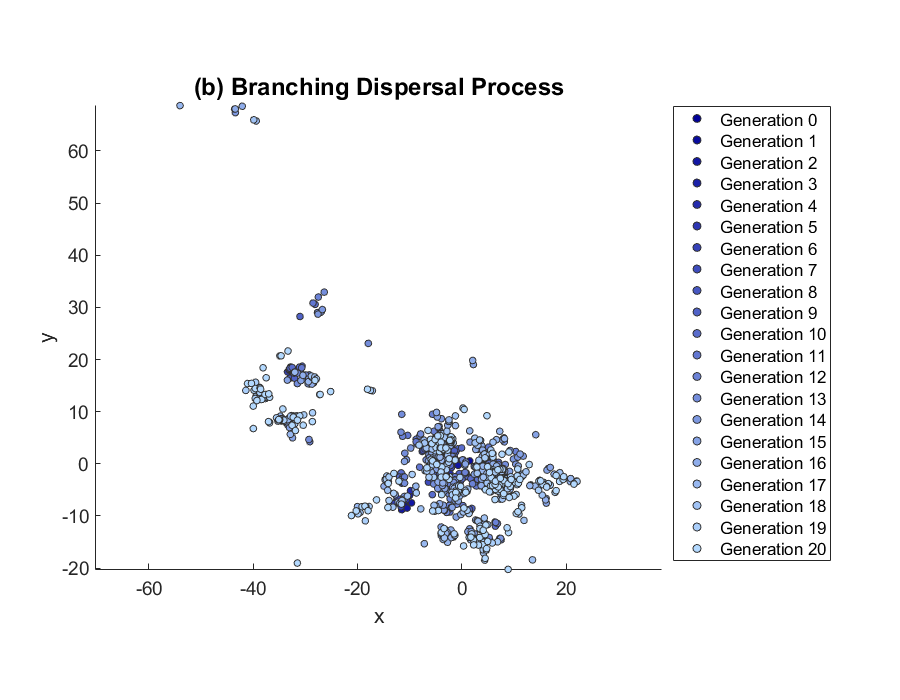}
    \caption{Two simulations of a spatial population process showing clumping and long-distance dispersal. The population is initialized at the origin with five individuals. Offspring numbers follow a Poisson distribution with mean 1.2, and offspring are dispersed radially according to $g(r)$ defined by \eqref{eq: g(r)}, with $H=0.35$, $a=b=0.5$, and direction uniformly distributed on $[0,2\pi]$. Each simulation is run for 20 generations beyond the initial population. }
    \label{fig: patchy invasions}
\end{figure}

The example of cheatgrass motivates Figure \eqref{fig: patchy invasions}, which shows two simulations of a population that starts at the origin, with 5 individuals. Each individual generates a Poisson number of children with mean 1.2. The offspring disperse according a stopped subdiffusive process with radial density \eqref{eq: g(r)} with $H=0.35$, $a=b=0.5$, and $\sigma=1$. Each  reproduces at the location where it stops. 
Differently from the simulations shown in Figure \eqref{fig: Figure 1}, where paths of the subdiffusive fBm were simulated, here we sample  from $g(r)$ in a three-step scheme: first we sample $\lambda\sim \text{Gamma}(0.5,0.5$), then the exponential stopping time $\tau\sim\mathcal{E}(\lambda)$, and finally we sample  from the Rayleigh distribution \eqref{eq: radial pdf}, $R\sim \text{Rayleigh}(\text{scale}=\tau^H)$, to obtain the radial distance, with the angle chosen uniformly from $[0,2\pi)$. We simulate  20 generations after the founding population and record their locations at each generation in a different shade of blue.

We observe the emergence of both short dispersal (the clumps), and long-distance dispersal (the long distance separation of the clumps), the latter arising from the heavy tail of the density $g(r)$. In real scenarios, heterogeneity in transport mechanisms may lead some offspring to travel substantially farther from their parents at each generation. Although such long-distance dispersers are relatively rare, they are sufficiently frequent to found several new patches moderately far from the parent population. Similar simulation outcomes are reported in (\cite{Lewis1997Variability, lewis2000modeling,lewis2016mathematics}). Their rather restrictive model employs two Laplace dispersal kernels with distinct means to represent separate movement scales assigned to different subsets of individuals. 

\section{Discussion}
The study of stopped random motions provides a mechanistic, yet simple, foundation for modeling dispersal in ecological systems. The pioneer work (\cite{BK53}) showed that isotropic Brownian motion stopped at an exponentially distributed random time produces a radial dispersal kernel with exponentially decaying tail, implying limited long-distance dispersal. The recent extension (\cite{gordillo2025parameter}) showed that introducing heterogeneity in stopping rates, for example, by drawing the exponential rate parameter $\lambda$ from a Gamma distribution, transforms the resulting kernel into one with heavy, power-law tail. Building upon this work, our present contribution establishes an analogous result even when the transport of particles follows a subdiffusive fBm, $H<1/2$, instead of a Brownian motion, $H=1/2$. We show, by analytic approximation and by simulation, that randomizing the stopping rate, $\lambda$, according to a Gamma distribution induces a heavy-tailed radial density $g(r) \sim C r^{-1-a/H}$ as $r\to \infty$, where the heavy tail in the radial distribution arises from the interplay between subdiffusive spreading and the heavy-tailed stopping times. 
This result offers a unified mechanistic explanation for the coexistence of strong local clumping and occasional long-distance dispersal events as observed in the movement of some invasive species. This construction integrates anomalous subdiffusive processes with realistic heterogeneity in individual settlement time distributions.

There are other models that explore related ideas. A popular model describing stratified dispersal in invasion biology is the ``$2Dt$ kernel" (\cite{Clark1999SeedDispersal,johnson2026modeling,lewis2016mathematics}), which assumes that the variance parameter of an underlying Gaussian dispersal process is randomized according to an inverse Gamma distribution. This mixture produces a Student's t distribution kernel of radial distances, which is heavy tailed. However, in the $2Dt$ model, movement of individuals is unbounded in principle (diffusion  continues indefinitely unless externally stopped). In contrast, in our model, each individual moves with fixed diffusivity until stopping according to a Gamma distributed rate $\lambda$. 
This random $\lambda$ introduces heterogeneity in the transport of individuals: some settle quickly near the source whereas others travel far, after a long time, before settling.
The $2Dt$ kernel may be a good fit to data in some instances, e.g. ``seed rain"  or  forests pests, and  is analytically convenient to capture variable ``mobility", that is, different wind gusts or vectors. However, our model has the advantage of being mechanistic for systems with a clear ``settlement" phase, e.g., seeds falling, larvae settling on substrate, animals stopping to forage. Our model addresses the point that individuals do not wander forever.

In the Discussion section of (\cite{johnson2026modeling}) there is a convincing argument pointing out that the $2Dt$, 2-parameter Student's-t distribution is a heavy-tailed (fat-tailed in their terminology) dispersal kernel which can allow for both short- and
long-distance events, i.e. stratified dispersal. This is followed by a thought-provoking overview of the relative merits of simple versus dynamically motivated stochastic models in this context.

Another interesting model  addressing clumping and long-distance dispersal is proposed in (\cite{koch2020}), where the authors highlight the Whittle–Matérn–Yasuda (WMY) kernel as a family of isotropic redistribution functions that generalizes several commonly used kernels in ecology.
The WMY kernel reflects a trade-off between a sharp central peak (clumping) and heavier tails (long-distance dispersal). The model comes from a mechanistic derivation of diffusion on a fractal domain with constant settling rate, which produces distributions that capture both local aggregation and occasional long jumps. The tail of the WMY kernel, however, produces tails heavier  than the Gaussian but still exponentially bounded. 

A number of variations of the problem studied here suggest themselves. For example, one could explore the case where the  stopping rate depends on space, or both space and time. The development of statistical methods to estimate $H,a,b$ and $\sigma$ from dispersal datasets would be necessary for applications. 

There are many examples of dynamical models involving a time rate where some event occurs followed by further evolution, such as diffusion. Such situations invite exploration of the case  where the rate is randomized, thereby introducing heterogeneity in the model.
\newline
\newline
\textbf{Data availability.} MATLAB codes used to generate the figures are available at \\https://github.com/LuisFGordillo/Codes
\bibliographystyle{plainnat}   
\bibliography{refs}

@book{biagini2008,
  title     = {Stochastic Calculus for Fractional Brownian Motion and Applications},
  author    = {Biagini, F. and Hu, Y. and {\O}ksendal, B. and Zhang, T.},
  year      = {2008},
  publisher = {Springer},
  address   = {London},
  series    = {Probability and Its Applications},
  isbn      = {978-1-85233-996-8},
  doi       = {10.1007/978-1-84628-797-8}
}

@article{BK53,
    author = "Broadbent, S.R. and Kendall, D.G.",
    title = "The random walk of \textit{Trichostrongylus retortaeformis}",
    journal = "Biometrics",
    volume = "9",
    number = "4",
    pages = "460--466",
    year = "1953",
}

@article{Clark1999SeedDispersal,
  author  = {Clark, J.S. and Silman, M. and Kern, R. and Macklin, E. and HilleRisLambers, J.},
  title   = {Seed Dispersal Near and Far: Patterns Across Temperate and Tropical Forests},
  journal = {Ecology},
  volume  = {80},
  number  = {5},
  pages   = {1475--1494},
  year    = {1999},
  doi     = {10.1890/0012-9658(1999)080[1475:SDNAFP]2.0.CO;2}
}

@book{johnson1994vol1,
  title     = {Continuous Univariate Distributions, Volume 1},
  author    = {Johnson, N. L. and Kotz, S. and Balakrishnan, N.},
  edition   = {2},
  year      = {1994},
  publisher = {Wiley},
  address   = {New York},
  isbn      = {9780471584957}
}

@article{johnson2026modeling,
  author = {Johnson, E.C. and Brush, M. and Lewis, M.A.},
  title = {Modeling stratified dispersal in forest pests: A case study of the mountain pine beetle in Alberta},
  journal = {Ecology},
  volume = {107},
  number = {2},
  pages = {e70305},
  year = {2026},
  month = {feb},
  doi = {10.1002/ecy.70305},
  publisher = {Wiley Online Library}
}

@article{koch2020,
  title   = {A Unifying Theory for Two-Dimensional Spatial Redistribution Kernels with Applications in Population Spread Modelling},
  author  = {Koch, D. and Lewis, M.A. and Lele, S.R.},
  journal = {Journal of the Royal Society Interface},
  volume  = {17},
  number  = {170},
  pages   = {20200434},
  year    = {2020},
  doi     = {10.1098/rsif.2020.0434}
}

@misc{gordillo2025parameter,
  title        = {Parameter variability can produce heavy tails in a model for the spatial distribution of settling organisms},
  author       = {Gordillo, L.F. and Greenwood, P.E.},
  year         = {2025},
  eprint       = {2509.16385},
  archivePrefix= {arXiv},
  primaryClass = {q-bio.PE},
  doi          = {10.48550/arXiv.2509.16385},
  url          = {https://arxiv.org/abs/2509.16385}
}

@misc{Molvar2024Cheatgrass,
  author       = {Molvar, E.M. and Rosentreter, R. and Mansfield, D. and Anderson, G.M.},
  title        = {Cheatgrass invasions: History, causes, consequences, and solutions},
  year         = {2024},
  institution  = {Western Watersheds Project},
  address      = {Hailey, ID},
  note         = {128 pp. Available online: https://westernwatersheds.org/wp-content/uploads/2024/02/Cheatgrass-Literature-Review-final.pdf}
}

@book{nair2022fundamentals,
  title     = {The Fundamentals of Heavy Tails: Properties, Emergence, and Estimation},
  author    = {Nair, J. and Wierman, A. and Zwart, B.},
  year      = {2022},
  publisher = {Cambridge University Press},
  series    = {Cambridge Series in Statistical and Probabilistic Mathematics},
  volume    = {53},
  isbn      = {9781316511732}
}

@article{Nathan2006LongDistance,
  author  = {Nathan, R.},
  title   = {Long-Distance Dispersal of Plants},
  journal = {Science},
  volume  = {313},
  number  = {5788},
  pages   = {786--788},
  year    = {2006},
  doi     = {10.1126/science.1124975}
}

@article{Nathan2008MovementEcology,
  author  = {Nathan, R. and Getz, W.M. and Revilla, E. and Holyoak, Ma. and Kadmon, R. and Saltz, D. and Smouse, P.E.},
  title   = {A movement ecology paradigm for unifying organismal movement research},
  journal = {Proceedings of the National Academy of Sciences},
  volume  = {105},
  number  = {49},
  pages   = {19052--19059},
  year    = {2008},
  doi     = {10.1073/pnas.0800375105}
}

@incollection{Lewis1997Variability,
  author    = {Lewis, M.A.},
  title     = {Variability, patchiness, and jump dispersal in the spread of an invading population},
  booktitle = {Spatial Ecology: The Role of Space in Population Dynamics and Interspecific Interactions},
  editor    = {Tilman, David and Kareiva, Peter},
  publisher = {Princeton University Press},
  address   = {Princeton, NJ},
  pages     = {46--69},
  year      = {1997}
}

@article{lewis2000modeling,
  title   = {Modeling and analysis of stochastic invasion processes},
  author  = {Lewis, M.A. and Pacala, S.W.},
  journal = {Journal of Mathematical Biology},
  volume  = {41},
  number  = {5},
  pages   = {387--429},
  year    = {2000},
  doi     = {10.1007/s002850000050},
  publisher = {Springer}
}

@book{lewis2016mathematics,
  title     = {The Mathematics Behind Biological Invasions},
  author    = {Lewis, M.A. and Petrovskii, S.V. and Potts, J.R.},
  year      = {2016},
  publisher = {Springer International Publishing},
  address   = {Cham},
  series    = {Interdisciplinary Applied Mathematics},
  volume    = {44},
  isbn      = {978-3-319-32042-7},
  doi       = {10.1007/978-3-319-32043-4}
}

@article{Shigesada1995Stratified,
  author  = {Shigesada, N. and Kawasaki, K. and Takeda, Y.},
  title   = {Modeling stratified diffusion in biological invasions},
  journal = {The American Naturalist},
  volume  = {146},
  number  = {2},
  pages   = {229--251},
  year    = {1995},
  doi     = {10.1086/285796}
}

@book{SamorodnitskyTaqqu1994,
  author    = {Samorodnitsky, G. and Taqqu, M.S.},
  title     = {Stable Non-Gaussian Random Processes: Stochastic Models with Infinite Variance},
  publisher = {Chapman \& Hall},
  address   = {New York},
  year      = {1994},
  isbn      = {0412051710}
}

@article{Trakhtenbrot2005LDD,
  author  = {Trakhtenbrot, A. and Nathan, R. and Perry, G. and Richardson, D.M.},
  title   = {The importance of long-distance dispersal in biodiversity conservation},
  journal = {Diversity and Distributions},
  volume  = {11},
  number  = {2},
  pages   = {173--181},
  year    = {2005},
  doi     = {10.1111/j.1366-9516.2005.00156.x}
}

@misc{Zouhar2003BromusTectorum,
  author       = {Zouhar, K.},
  title        = {Bromus tectorum},
  year         = {2003},
  howpublished = {Fire Effects Information System (FEIS), U.S. Department of Agriculture, Forest Service, Rocky Mountain Research Station, Fire Sciences Laboratory},
  note         = {Available online: https://www.fs.usda.gov/database/feis/plants/graminoid/brotec/all.html},
}
\end{document}